\documentclass[prl,onecolumn,a4paper,superscriptaddress]{revtex4}

\usepackage{amsmath}
\usepackage{graphicx}
\usepackage{sistyle}
\usepackage{todonotes}
\usepackage{gensymb}
\usepackage{bbold}
\usepackage{bm}
\usepackage{natbib}
\usepackage{amsfonts}
\usepackage{amssymb}
\usepackage{latexsym}
\usepackage{fancyhdr}
\usepackage[bbgreekl]{mathbbol}
\usepackage{epsfig}
\usepackage{color}
\usepackage{textcomp}

\def\ket#1{\mathinner{|{#1}\rangle}}
\def\mean#1{\langle #1 \rangle}

\def\text#1{\textrm{#1}}
\DeclareMathOperator{\Tr}{Tr}
\newcommand{\pw}[1]{ \cdot 10^{#1}}
\def\prjct#1{\mathinner{|{#1}\rangle}\!\!\mathinner{\langle{#1}|}}
\def\op#1{\hat{#1}}
\def\commut#1#2{\left\lbrack #1,#2 \right\rbrack}
\renewcommand{\ao}[1][]{%
	\ifthenelse{\equal{#1}{}}{\ensuremath{\op{a}}}{\ensuremath{\op{#1}}}%
}

\newcommand{\co}[1][]{%
	\ifthenelse{\equal{#1}{}}{\ensuremath{\op{a}^{\dagger}}}{\ensuremath{\op{#1}^{\dagger}}}%
}

\def\dissip#1{\mathcal{D}\left[#1\right]}
\def\Dissip#1#2{\mathcal{D}\left[#1\right]#2 = #1#2#1^{\dagger}-\dfrac{1}{2}\left(#1^{\dagger}#1#2+ #2#1^{\dagger}#1 \right)}
\usepackage{xparse}
\NewDocumentCommand\Lindbladian{mg}{%
	\ensuremath{\mathcal{L}\IfNoValueTF{#2}{}{_{#2}}\left(#1\right)}%
}
\newcommand{\modif}[1]{{\color{black}#1}}

\begin{document}

\title{Two-Color Pump-Probe Measurement of Photonic Quantum Correlations Mediated by a Single Phonon}

\author{Mitchell D. Anderson$^\dagger$}
\author{Santiago Tarrago~Velez$^\dagger$}
\author{Kilian Seibold}
\author{Hugo Flayac}
\author{Vincenzo Savona}
\affiliation{Institute of Physics, Ecole Polytechnique F\'{e}d\'{e}rale de Lausanne (EPFL), CH-1015 Lausanne, Switzerland}
\author{Nicolas Sangouard}
\affiliation{Departement Physik, Universit\"{a}t Basel, Klingelbergstrasse 82, CH-4056 Basel, Switzerland}
\author{Christophe Galland}
\affiliation{Institute of Physics, Ecole Polytechnique F\'{e}d\'{e}rale de Lausanne (EPFL), CH-1015 Lausanne, Switzerland}

\date{\today}

\begin{abstract}

We propose and demonstrate a versatile technique to measure the lifetime of the one-phonon Fock state using two-color pump-probe Raman scattering and spectrally-resolved, time-correlated photon counting.
Following pulsed laser excitation, the $n=1$ phonon Fock state is probabilistically prepared by projective measurement of a single Stokes photon.
The detection of an anti-Stokes photon generated by a second, time-delayed laser pulse probes the phonon population with sub-picosecond time resolution. We observe strongly non-classical Stokes--anti-Stokes correlations, whose decay maps the single phonon dynamics. 
Our scheme can be applied to any Raman-active vibrational mode. It can be modified to measure the lifetime of $n \geq 1$ Fock states or the phonon quantum coherences through the preparation and detection of two-mode entangled vibrational states.

\end{abstract}

\maketitle

\paragraph{Introduction ---}

Phonons, the quantized excitations of internal vibrational modes in crystals and molecules, span a broad frequency range up to $\sim 100$ THz. At these high frequencies, thermal occupancy at room temperature is much less than one, so that quantum effects are readily observable. For example, creation and annihilation of a single phonon within one short laser pulse produces non-classically correlated Stokes--anti-Stokes (SaS) photon pairs \cite{klyshko1977,parra-murillo2016,schmidt2016}, as observed in pulsed Raman scattering from a diamond crystal \cite{kasperczyk2015}, liquid water \cite{kasperczyk2016} and other molecular species \cite{saraiva2017}. 
With the advent of quantum optomechanics, the quantisation of lower frequency (MHz to GHz) mechanical oscillations was also evidenced in several experiments using phase sentive detection \cite{purdy2017,sudhir2017} and photon counting \cite{riedinger2016,hong2017a}.  
Finally, in a series of recent experiments, Raman-active phonon modes in pure diamond \cite{lee2012,england2013,england2015,hou2016,fisher2016a,fisher2017} and gaseous hydrogen \cite{bustard2013,bustard2015,bustard2016} have been used to store and process classical and quantum information on picoseconds time scales at room-temperature. 
Developing versatile schemes and techniques to address non-classical phonon states in bulk and nanoscale systems is thus a promising research direction to improve our understanding of quantum effects occurring at ambient conditions and leverage them for quantum technologies.

\begin{figure}[t!]
\centering
\includegraphics[width=8cm]{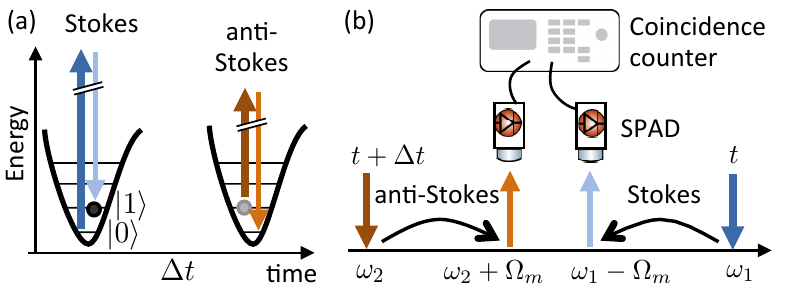}
\caption{\label{fig:scheme} (a) Concept of the experiment: The first (write) laser pulse probabilistically prepares the phonon in state $\ket{n=1}$ upon detection of a Stokes photon. The second (read) pulse, after a time delay $\Delta t$, converts this phonon into an anti-Stokes photon. (b) Schematics of the photon counting in the frequency domain (SPAD: single-photon avalanche photodiode). The choice of $\omega_1$ and $\omega_2$ is completely free, as long as the Stokes and anti-Stokes photons can be efficiently isolated by spectral filtering. 
}\end{figure}

In this Letter, we present a new scheme to measure the creation and annihilation of a single phonon Fock state with sub-picosecond time resolution (Fig.~\ref{fig:scheme}), which can be applied on any Raman-active high frequency mode, such as ubiquitously found in organic materials. \modif{Our scheme is conceptually similar to the one recently applied to an optomechanical cavity with a GHz mechanical oscillator \cite{riedinger2016}, although we don't use any optical cavity and measure the dynamics on time scales that are 5 to 6 orders of magnitude shorter.}
We use diamond in a proof-of-principle experiment (phonon frequency $\Omega_m/2\pi=39.9$~THz), but in contrast to Refs. \cite{lee2012,england2013,england2015,hou2016,fisher2016a,fisher2017,bustard2013,bustard2015,bustard2016}, our scheme does not rely on the polarization selection rule of Raman scattering to temporally distinguish between photons. It is therefore not restricted to diamond and can be applied to low-dimensional structures and molecules as well, in the solid, liquid or gaseous phase. 
Indeed, we use two-color excitation and spectral multiplexing to distinguish the photons from the \textit{write} and \textit{read} steps. In the \textit{write} step, a laser pulse centered at frequency $\omega_1$ leads to Stokes scattering with low probability $p_S \ll 1$ (two-phonon generation occurs with probability $\propto p_S^2 \ll p_S$). Detection of a Stokes (S) photon at frequency $\omega_1-\Omega_m$ projects the phonon onto the Fock state $\ket{n=1}$. In the \textit{read} step, after a controllable time delay $\Delta t$, a second synchronised pulse centered at a different frequency $\omega_2$ is used to probe the population of the conditional phonon Fock state by detection of an anti-Stokes (aS) photon at frequency $\omega_2+\Omega_m$. 
\modif{The value of the second-order cross-correlation $g^{(2)}_{S,aS}(\Delta t)$ between the S and aS photons witnesses the non-classical nature of the two-photon state produced by the exchange of a single phonon \cite{glauber1963a} (see \cite{SM} Sec.~1, for a more general discussion).} The dynamics of this non-classical SaS correlation can be tracked by scanning $\Delta t$, revealing the single-phonon lifetime.

\paragraph{Experimental Setup ---} 

\begin{figure}[h]
\centering
\includegraphics[width=7.5cm]{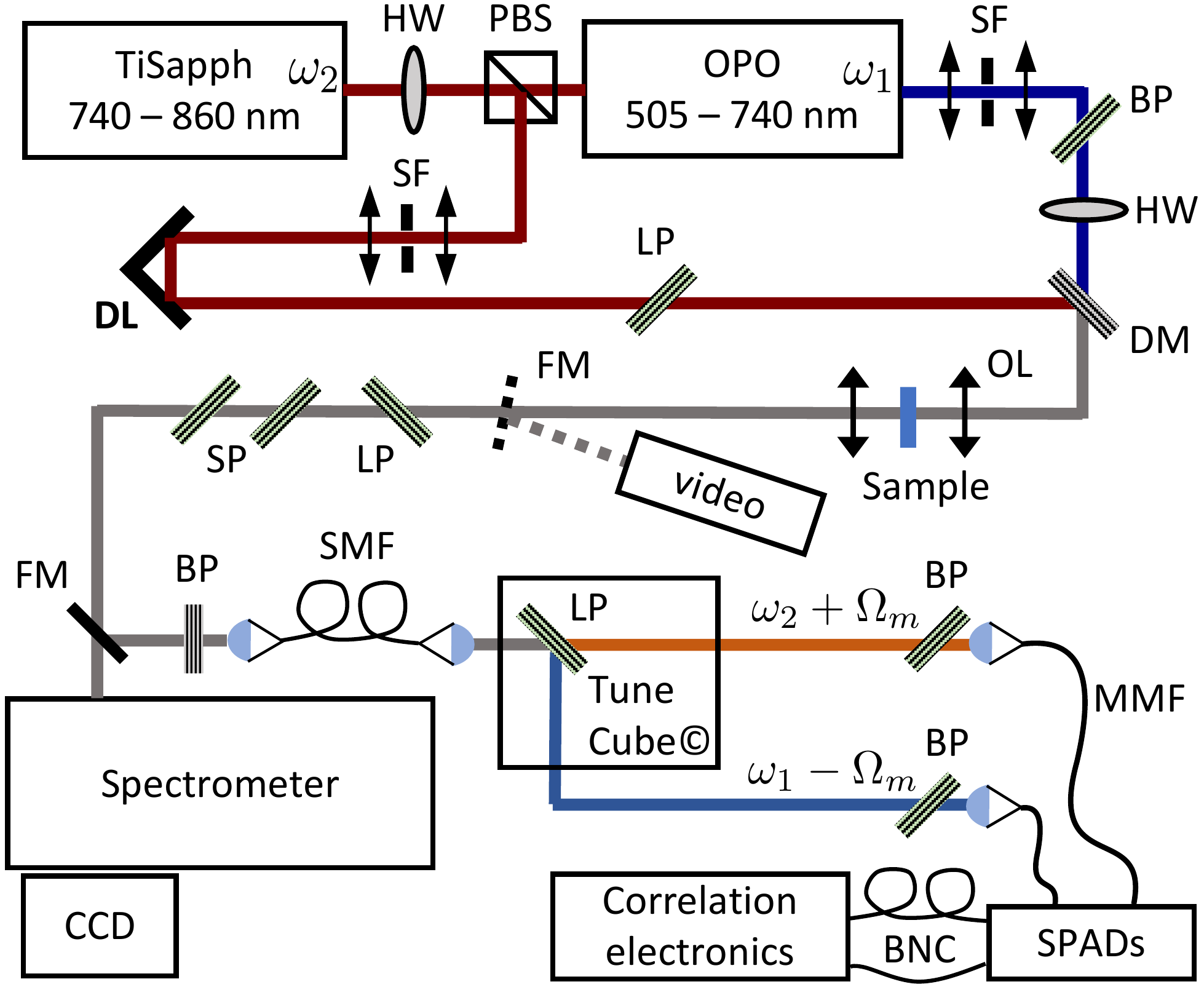}
\caption{\label{fig:setup} Schematic drawing of the experimental setup. HW: half-wave plate; PBS: polarizing beam-splitter; SF: spatial filter; SP/LP/BP: short/long/band-pass filters (tunable filters are highlighted in green); DL: delay line; DM: Dichroic mirror; OL: objective lens (numerical aperture = 0.8); SMF: single-mode fiber (HP780, Thorlabs); MMF: graded index multimode fiber (100~$\mu$m core, NA 0.29, OZ Optics); FM: flip mirror. A video camera is used to overlap the beams.
}\end{figure}

Our experimental setup is depicted in Fig.~\ref{fig:setup}. 
The two synchronized femtosecond pulse trains are generated by a Ti:Sapph oscillator (Tsunami, Spectra Physics, 80~MHz repetition rate) and a frequency-doubled optical parametric oscillator (OPO-X fs, APE Berlin). 
We can independently tune the Ti:Sapph wavelength between 740 and 860~nm and the OPO wavelength between 505 and 740 \cite{APE}. 
The OPO generates the {write} pulse, while the Ti:Sapph is sent through a delay line to provide the {read} pulse. 
The {read} and {write} pulses are combined at a dichroic mirror before being focused on a synthetic diamond crystal ($\sim 300~\mu$m thick) cut along the 1-0-0 crystal axis. 
We use tunable interference filters (highlighted in green in Fig.~\ref{fig:setup}) to block the spectral components of the excitation pulses that overlap with the detection window. The sample is studied in transmission with a pair of objective lenses in order to fulfill momentum conversation in the exchange of the same phonon in the read and write scattering processes.
After the sample, we block most of the laser light with a combination of tunable short and long pass filters, and send the signal either to a spectrometer equipped with a cooled CCD array or to a single mode fiber, which selects a single spatial mode of the photons. Since momentum is conserved during Raman scattering, this allows us to probe a well-defined phonon spatial mode in the bulk crystal. 
After the fiber, light is collimated and sent to a tunable dichroic mirror (TuneCube, AHF analysentechnik AG), which allows us to separate the S and aS photons, depending on their wavelengths, by rotating a tunable filter (here a long-pass). 
The separated signals are further spectrally filtered before impinging on fiber-coupled single photon avalanche photodiodes (SPADs) connected to a coincidence counter.

\paragraph{Results ---} 

In each experiment, we begin by tuning the Ti:Sapph and OPO to center frequencies $\tilde{\omega}_2$ and $\tilde{\omega}_1$ so that $\tilde{\omega}_1-\tilde{\omega}_2=\Omega_m$. 
When the two pulses overlap, both spatially and temporally, strong coherent anti-Stokes Raman scattering (CARS) at frequency $\tilde{\omega_1}+\Omega_m$ is generated. 
We use this signal to find the zero time delay and optimize the spatial overlap of the two excitation beams.

\begin{figure}[t!]
\centering
\includegraphics[width=8cm]{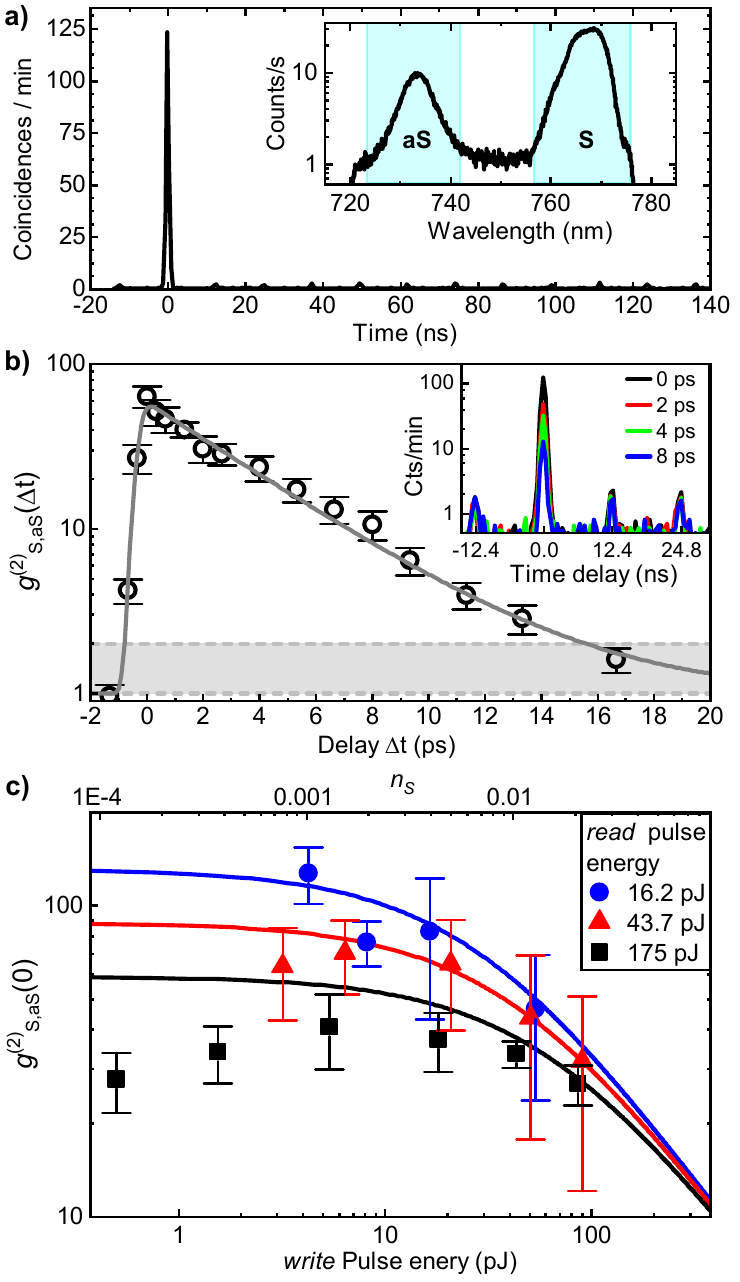}
\caption{\label{fig:g2SAS} (a) S--aS coincidence histogram (20 min integration). The S (resp. aS) count rates were $< 2.5\pw{4}$~Hz (resp. $< 8\pw{2}$~Hz), with a dark count rate $< 5\pw{2}$~Hz. The OPO (resp. TiSapph) powers just after the sample were $\sim 1.5$~mW (resp. $\sim 3.5$~mW).
Inset: representative Raman spectrum; the pass-bands of the filters placed before the S and aS detectors are overlaid in blue. 
(b) \textit{Write-read} delay dependence of the measured S--aS correlation (open circles) and fit (dashed line) using an exponential decay (time constant = $3.9\pm 0.7$~ps) convoluted with the instrument response (Gaussian with standard deviation $\sigma=223$~fs, see \cite{SM}). The gray area marks the classical bounds. Inset: coincidence histograms at different delays. (c) Measured SaS correlations at zero delay vs. average power in the \textit{write} beam (symbols) for different powers in the \textit{readout} beam. 
\modif{Solid lines are the result of the simplified analytical model presented in \cite{SM} Sec.~4. It has two free parameters: the overall collection and detection efficiency $\eta$ (probability of registering a S or aS photon emitted in the right spatial mode) and a factor $\alpha$ accounting for the non-ideal phonon-photon conversion in the read pulse, which we relate to imperfect mode overlap. The best fit to the data is obtained with $\eta=0.07$ and $\alpha=0.3$.}
}\end{figure}

The center frequency of the {write} pulse is then tuned so that the S and aS peaks are spectrally separated from the {read} and {write} pulses.
As a first demonstration, the central wavelengths of the {write} and {read} pulses are set to 696 nm and 810 nm, respectively.
This results in S photons at 767 nm (1.619 eV) and aS photons at 732 nm (1.695 eV), as seen on the Raman spectrum of Fig. \ref{fig:g2SAS}a, inset.
Figure \ref{fig:g2SAS}a presents the coincidence histogram obtained at zero {write-read} delay in this configuration. 
The $t=0$~ns peak corresponds to events where one photon is detected in each channel within the same {write-read} pulse sequence. 
Since the delay between two repetitions (12.5~ns) is three to four orders of magnitude longer than the phonon lifetime, the side peaks are due to uncorrelated photons (``accidental" coincidences). The number of coincidences in the central peak, divided by the average number of coincidences in the side peaks, is a measure of $g_{S,aS}^{(2)}(0)$, the normalized second-order cross-correlation function between the S photons produced in the {write} pulse and the aS photons produced in the {read} pulse \cite{riedinger2016,galland2014}.

The Cauchy-Schwartz inequality sets an upper bound on the possible value of the cross-correlation for classical fields $g_{S,aS}^{(2)}(0) \leqslant \left( g_{S,S}^{(2)}(0) g_{aS,aS}^{(2)}(0)\right)^{1/2}$, where the terms on the RHS are the second-order auto-correlation functions of the S and aS fields \cite{reid1986,kuzmich2003}. We expect that $g_{S,S}^{(2)}(0)=2$ since the spontaneously emitted S photons follow the same (thermal) statistics as in parametric down-conversion below threshold \modif{(See \cite{SM}, Sec.~1-2)}. However, this is true only in the single-mode situation, as the the Stokes auto-correlation function falls as $1+1/N$ with $N$ the number of phonon and photon modes (see \cite{SM}, Sec.~1-2). 
We therefore used a 50/50 beam splitting fiber to measure the auto-correlation of the S channel and found $g_{S,S}^{(2)}(0)=2 \pm 0.1$ (see Fig.~S1), confirming that our experiment measures the state of a single phonon mode. Although the count rate on the aS detector was not sufficient to measure precisely $g_{aS,aS}^{(2)}(0)$, we cannot think of any reason why it should be larger than $2$  since the aS photons should carry the thermal statistics of the phonon mode. In summary, the classical bound to our measurement is $g_{S,aS}^{(2)}(0) \leqslant 2$.
The measured value  $g_{S,aS}^{(2)}(0) = 63.4 \pm 9.7$ in Fig.~\ref{fig:g2SAS}a,b thus \modif{violates the Cauchy-Schwartz inequality by 6 standard deviations} and is a proof of quantum correlations between the S and aS photons, mediated by the exchange of a single phonon. 

We then repeat the coincidence measurement for many different positions of the delay line and obtain the time dependent correlation function $g_{S,aS}^{(2)}(\Delta t)$ (Fig.~\ref{fig:g2SAS}b). The correlations decay with a $1/e$ time constant of $3.9\pm 0.7$~ps  (bounds for 95\% confidence), in agreement with the literature values of the optical phonon lifetime in diamond \cite{lee2012,liu2000}. This demonstrates that we are able to measure the lifetime of a phonon Fock state by following the decay of non-classical S--aS correlations. 

In order to understand what determines the precise value of $g_{S,aS}^{(2)}$ and what limits the achievable degree of non-classical correlations,
we study the dependence of the zero-delay correlation $g_{S,aS}^{(2)}(0)$ on the powers in the {write} ($P_w$) and {read} ($P_r$) beams (Fig.~\ref{fig:g2SAS}c) and compare the results to an analytically soluble quantum model of parametrically coupled photon-phonon modes at zero temperature (\cite{SM}, Sec.~2). In direct analogy with the physics of photon pairs produced by parametric down-conversion \modif{(see also \cite{SM}, Sec. 1)} \cite{sekatski2012}, we find that $g_{S,aS}^{(2)}(0)$ decreases as $1/n_S \propto 1/P_w$ where $n_S$ is the average S photon number produced by the {write} pulse. This can be understood as the consequence of the growing probability of exciting the $\ket{n=2}$ phonon Fock state at higher power compounded with the fact that our detectors cannot resolve the photon number.

Interestingly, at low $P_w$ the correlation saturates at a value that depends on the power in the {read} pulse $P_r$. We can explain this behavior by the noise generated in the {read} pulse, which has three components. (i) The thermal phonons (thermal occupancy $n_{th} < 2 \pw{-3}$) are responsible for uncorrelated aS emission. If this were the only source of noise, then $g_{S,aS}^{(2)}(0) \rightarrow 1/n_{th}$ at low {write} powers $P_w$, irrespective of the {read} power $P_r$. (ii) Yet, another intrinsic noise source related to the Raman process is SaS pair emission in the {read} pulse \cite{parra-murillo2016}, which scales quadratically with $P_r$. (iii) Finally, we identified spontaneous four-wave mixing \cite{SM} as another source of uncorrelated counts on the aS detector in the {read} pulse. In the SM, Sec.~3, we present a more complete quantum model in which the effects (i-iii) are accounted for. Its dynamics is solved numerically at non-zero temperature and the results reproduce qualitatively well our observations without fitting parameters, suggesting that we have a comprehensive understanding of the noise sources limiting the value of $g_{S,aS}^{(2)}(0)$. 

The simplified model used to fit the data in Fig.~\ref{fig:g2SAS}c can also be used to compute the expected second-order auto-correlation of the aS photons $g_{aS,aS}^{(2,cond)}(0)$ conditional on the detection of a S photon in the {write} pulse, as would be measured to characterize heralded single-photons \cite{galland2014} (see \cite{SM}). We find values of $g_{aS,aS}^{(2,cond)}(0)$ well below $0.1$ for the parameters corresponding to most data points of Fig.~\ref{fig:g2SAS}c, demonstrating that our experiment indeed probes the dynamics of the $\ket{n=1}$ phonon Fock state, with negligible contribution of $n \geqslant 2$ eigenstates.

A route to increase the measured quantum correlation toward the $1/n_{th}$ thermal limit is the use of a cavity in the resolved sideband regime to select only S and aS processes in the {write} and {read} pulses, respectively \cite{galland2014}. This is where the broad tunability of our setup would become particularly relevant. 

\begin{figure}[h]
\centering
\includegraphics[width=7.5cm]{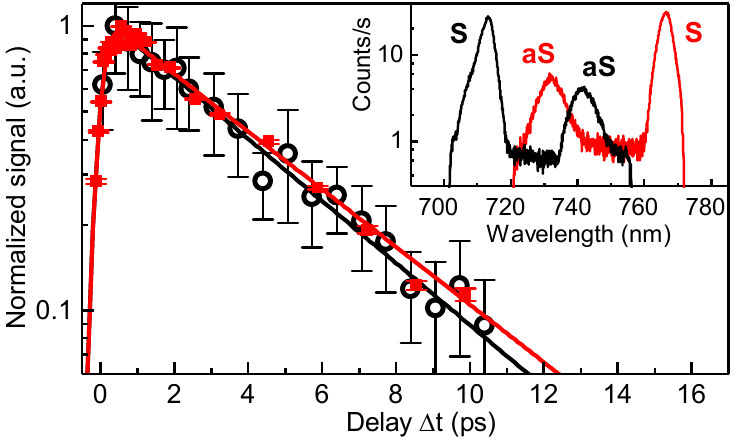}
\caption{\label{fig:config2} a) Decay of the normalized SaS correlation $(g_{S,aS}^{(2)}(\Delta t)-1)/(g_{S,aS}^{(2)}(0)-1)$) (symbols) as a function of {write--read} delay $\Delta t$ for two configurations of laser wavelengths (red and black colors), yielding S and aS peaks as shown in the inset. The fitted exponential decays (solid lines) have time constants  $4.2 \pm 0.5$~ps (red line) and $4.0 \pm 0.3$~ps (black line).
}\end{figure}
 
Indeed, we show in Fig.~\ref{fig:config2} that we can perform this measurement using different configurations of the excitation/detection wavelengths accessible with our instruments. For example, the {write} and {read} pulses were also set at 650 nm and 821 nm, respectively, yielding S and aS photons at 712 nm (1.74 eV) and 740 nm (1.67 eV) (see Fig. \ref{fig:config2}, inset).  
Although the absolute value of $g_{S,aS}^{(2)}(0)$ depends on the laser powers, on the quality of alignment and beams' overlap, on the amount of cross-talk between the S and aS channel and on the amount of background emission, after normalization both data sets accurately track the phonon dynamics. This demonstrates the broad tunability of our setup (limited here by the available filters) and the robustness of our technique.

\paragraph{Conclusion ---} 
Our scheme constitutes a broadly applicable technique for the time-resolved measurement of quantum correlations mediated by high frequency vibrational modes, which can be observed even at room-temperature due to their vanishing thermal occupancy. As we verified by rotating the linear polarization of the {write} beam, our scheme is polarization insensitive, so that it can be applied to any Raman-active mode. It is well suited to study quantum dynamics in individual nanosystems -- in principle down to a single molecule \cite{yampolsky2014}. 
As shown in Ref.~\cite{lee2012,england2013,england2015,hou2016,fisher2016a,fisher2017,bustard2013,bustard2015,bustard2016}, Raman-active phonons are potential candidates for room-temperature quantum information processing. Our scheme extends the feasibility of this approach to a much broader range of material systems, which can be optimized for coupling efficiency and longer phonon lifetime. 

The wide tunability of our setup will allow to leverage the resonant enhancement provided by electronic transitions or nanocavities, 
while spectral multiplexing and photon counting make it possible to measure cross-correlations between different normal modes (by triggering the start and stop detectors with two different Raman lines), thereby probing inter-mode coupling dynamics. Moreover, by triggering the coincidence counter upon multi-photon detection in the {write} step (using spatial \cite{divochiy2008,dauler2009} or temporal \cite{fitch2003} multiplexing or a direct photon number resolving detector \cite{kim1999,miller2003}), our technique would probe the dynamics of higher vibrational Fock states ($n>1$) \cite{wang2008}. The probabilistic nature of the scheme, however, means that the rate of successful events will drop exponentially with $n$.
Finally, this work constitutes the basis for more advanced measurement schemes where phonon coherences are measured using vibrational two-mode entangled 
states \cite{flayac2014} and photon-phonon entangled states \cite{vivoli2016}.
This could  lead to new ways of studying quantum phenomena in organic systems, which play essential roles in photochemistry and possibly in some biological reactions \cite{wilde2010,oreilly2014}.

\paragraph{Acknowledgements ---} We thank Pascal Gallo and Niels Quanck for providing diamond crystals; Mark Kasperczyk for helpful discussions regarding the experiment, Philippe Roelli, Nils Kipfer and Tianqi Zhu for assistance in the laboratory and Vivishek Sudhir for insightful discussion of the results. This work was made possible by the Swiss National Science Foundation (SNSF), through the grants number PP00P2-170684 and PP00P2-150579.\\
$^\dagger$ M.D.A. and S.T.V. contributed equally to this work.

\bibliography{SaS_correlations_biblio}

\pagebreak
\begin{center}
\LARGE{Supplementary Material}
\end{center}


\modif{\section{Primer on the Cauchy-Schwartz inequality for correlated photon pairs}\label{sec:primer}

In this section, we introduce a few basic concepts essential to the understanding of the main text when discussing the violation of the Cauchy-Schwartz inequality (CSI) related to the value of the Stokes--anti-Stokes cross-correlation $g^{(2)}_{S,aS}\doteq g^{(2)}_{S,aS}(0)$. 
This section is based on existing literature (in particular \cite{sekatski2012}) and the reader is referred to it for more detailed discussions. 

\vskip 3mm

\paragraph{Witnessing non-classical correlations via the violation of the CSI ---}

The density matrix of a two-mode state $\rho_{a b}$ can be expressed in the $P$-representation as
\begin{equation}
\rho_{ab}=\int d^2\alpha\, d^2\beta P(\alpha,\beta) \prjct{\alpha, \beta},
\end{equation}
where $\alpha$ and $\beta$ are complex number and $\ket{\alpha}$, $\ket{\beta}$ are coherent states for $a$ and $b$ respectively. The state $\rho_{ab}$ is said to be classical if the quasi-probability distribution $ P(\alpha,\beta)$ is non-negative \cite{reid1986}. In other words, bipartite classical fields can be described by a statistical mixture of coherent states and their statistics can be reproduced by modulating the phase and amplitude of laser fields. In this case, the correlation functions associated to modes $a$ and $b$ fulfill
\begin{equation}\label{R_factor}
R=\frac{\mean{a^\dag b^\dag b \,a}^2}{\mean{{a^{\dag 2}} {a}^2}\mean{b^{\dag^2} b^2}} = \frac{\left( g^{(2)}_{a,b} \right)^2}{g^{(2)}_{a,a}g^{(2)}_{b,b}}\leq 1.
\end{equation}
This inequality is called the CSI and is saturated for pure coherent states ($R=1$) \cite{reid1986}. The violation of the CSI is thus a sufficient condition for non-classicallity of a two-mode state. In contrast, for a bi-photon number state $|11\rangle$, $R \rightarrow \infty$. 

\vskip 3mm

\paragraph{Relationship between parametric down-conversion and our experiment --- } 

Experimentally, photon pairs are typically produced by spontaneous parametric down-conversion, a nonlinear process in which a single photon from the pump is annihilated while two photons of lower energies are created in each mode $a$ and $b$ so that energy and momentum are conserved. Neglecting all possible sources of noise (such as thermal photons), the resulting photons (if suitably filtered) are in a two-mode squeezed state, for which $R=\frac{1}{4}(1+\frac{1}{p})^2$ (see e.g.~\cite{kuzmich2003}), where $p$ is the probability of pair emission, which depends linearly on the pump power for small $p$ (regime of spontaneous down-conversion), so that ideally $R \rightarrow \infty$ when $p \rightarrow 0$.
In our experiment, modes $a$ and $b$ correspond to the Stokes and anti-Stokes photons produced in the first and second pulses, respectively. They are correlated via the creation and annihilation of a single phonon in a specific phonon mode $c$. As we detail in the following Section~\ref{sec:analytic}, we can therefore establish a rigorous mapping between the two-mode squeezed state generated in parametric down-conversion and the two-mode Stokes--anti-Stokes state produced in our experiment. An important difference though us that non-classical correlations existing between Stokes and anti-Stokes are mediated by the phonon mode and persist over several picoseconds. Moreover, Stokes and anti-Stokes photons can be produced at arbitrary frequencies by tuning the two lasers. The phonon thus acts as a short-term quantum memory. 

\vskip 3mm

\paragraph{$g^{(2)}_{S,aS} > 2$ as a sufficient condition to violate the CSI ---} 

An upper bound of the denominator of $R$ can be generally established for photons produced in a nonlinear process such as Raman scattering. In the regime of spontaneous Raman scattering, the distribution of either Stokes or anti-Stokes photon number is stochastic (thermal statistic), so that $g^{(2)}_{S,S} \leqslant 2$ and $g^{(2)}_{aS,aS} \leqslant 2$. Therefore, a sufficient condition for $R > 1$ is that $g^{(2)}_{S,aS} > 2$. In our experiment, we could indeed verify directly that $g^{(2)}_{S,S} \leqslant 2$ (see Fig.~\ref{fig:Nmode}). It was difficult to measure the value of $g^{(2)}_{aS,aS}$ due to the low count rate, but no known mechanism can yield $g^{(2)}_{aS,aS} > 2$ in our situation.

\vskip 3mm

\paragraph{Maximum violation of the CSI in a quantum model ---} As we mentioned above, the bi-photon number state $|11\rangle$ maximally violates the CSI. It is however a limiting case of what is experimentally feasible. In our experiment, the value of $R$ is fundamentally limited by thermal noise in the phonon number (which mediates the photon number correlations), which we quantify by the thermal phonon occupancy $n_{th}$ ($n_{th}\approx 1.7 \pw{-3}$ at room temperature for the phonon frequency in our experiment). To understand why, let's examine the probability of emitting an anti-Stokes photon in the read pulse $p_{aS}$. For a given read pulse energy, $p_{aS}$ is directly proportional to $n_{c}$, the phonon number ($p_{aS} \equiv \beta n_c$). 

In the regime of very low Stokes emission probability $p$ in the write step, i.e. when the average Stokes photon number $\bar{n}_S = p < n_{th}$, such that the average phonon population $\bar{n}_c \approx n_{th}$, we have in the read step $p_{aS} \approx \beta n_{th}$. Yet, conditional on the detection of a Stokes photon in the write step, i.e. when a phonon is created, then immediately afterward $n_c \approx 1$ and $p_{aS|S} = \beta$. Since by definition $g^{(2)}_{S,aS}=\frac{P_{S \cap aS}}{P_{S} P_{aS}}=\frac{P_S P_{aS|S}}{P_{S} P_{aS}}=\frac{P_{aS|S}}{P_{aS}}$ we deduce from these simple considerations that $g^{(2)}_{S,aS} \leqslant 1/n_{th} \sim 600$. This result was confirmed by rigorous quantum calculations (Sec.~\ref{sec:analytic} and \ref{sec:numeric}).

In the regime of higher Stokes emission probability $n_{th} < p < 1$,  we show in Sec.~\ref{sec:analytic} that $g^{(2)}_{S,aS}$ decreases approximately as $1/p$, similarly to parametric down-conversion sources. 

In the main text and in Sec.~\ref{sec:back}, we show that other sources of technical and intrinsic noise further reduce the maximally achievable $g^{(2)}_{S,aS}$.

\vskip 3mm

\paragraph{Relationship between CSI violation and Bell-type inequality violation --- }

We briefly mention how to relate the CSI to Bell-type inequalities; see for example a similar discussion in Ref.~\cite{farrera2016a}. Let us stress first that experimental violation of a Bell inequality witnesses a much stronger form of non-classicality, related to entanglement, than CSI violation does (see for example the discussion in \cite{vivoli2016}). This being said, if our experiment were modified to measure photon-photon entanglement mediated by the phonon, for example using polarization or time-bin entanglement between photon and phonon, the maximal visibility of the two-photon interference fringes in a Bell-type measurement would be $V_{max}=\frac{g^{(2)}_{S,aS}-1}{g^{(2)}_{S,aS}+1}$. Therefore, when $g^{(2)}_{S,aS} > \frac{1+1/\sqrt{2}}{1-1/\sqrt{2}} \approx 5.83$ we have $V_{max}>1/\sqrt{2}$, in principle enabling a violation of the Clauser-Horne-Shimony-Holt (CHSH) Bell inequality.

}


\section{Analytical multi-mode quantum model with noise}\label{sec:analytic}

Here, we present an analytical multi-mode model, including non-ideal detectors, to compute Stokes (S) and anti-Stokes (aS) auto- and cross-correlations, based on the formalism used to describe parametric down-conversion photon sources (details can be found in Ref.~\cite{sekatski2012}). We will first show that the auto-correlation measurements performed on the Stokes mode are in perfect agreement with the prediction of the model when a single phonon mode is excited. Then, we use the model to reproduce the behavior of the SaS cross-correlation as a function of the powers in the {write} and {read} pulses. Finally, we present the results that would be obtained by performing a conditional auto-correlation measurement of the aS photons. These results provide a very good indication that single phonon Fock states are being addressed in the experiment.

Let us mention that the proposed model is valid in the limit of zero-temperature and of short time delay compared to the phonon lifetime (the decay of the phonon mode is not included), so we use it to study the value of the measured cross-correlations at zero delay. We account for the thermal phonons through the noise they produce in the anti-Stokes mode in the {readout} step. Since the total noise on the aS detector can be measured directly by switching off the write pulse, we use its value as input in the model, so that thermal noise is implicitly included. 

\vspace{3mm}
\paragraph{Model ---}

We introduce the annihilation operators $c_n$ for the $N$ phonon modes, and the operators $a_n$ for the corresponding Stokes photon modes. With this labeling, the joint photon-phonon state immediately after the first pulse is described by the density matrix
\begin{equation}
\label{stateN}
\rho_{ac}^N=(1-\bar p)^N e^{\sqrt{\bar p} \sum_{n=1}^{N} a_n^{\dag}c_n^{\dag}} |0\rangle\langle0|  e^{\sqrt{\bar p} \sum_{n=1}^{N} a_n c_n} 
\end{equation}
where the emission probability $\bar p$ is such that the average photon number of the N modes $a_n$ (or $c_n$) is given by $N\frac{\bar p}{1- \bar p}.$

As explained in Ref.~\cite{sekatski2012}, a realistic multimode single-photon detector can be modeled by the operator:
\begin{equation} \label{deta}
D_{\sum_n a_n}(\eta_a,q_a)\doteq D_a^N = \mathbf{1}-(1-q_{a})(1-\eta_a)^{\sum_n a_n^{\dag}a_n}
\end{equation} 
where $\eta_a$ is the detection efficiency (same for all modes for simplicity) and $q_a$ the dark count probability. 
More generally, one can include in $q_a$ all detection events that are not due to the Stokes process; this can be leakage of laser light or photons created by another nonlinear process (Four Wave Mixing) or fluorescence. These photons are typically synchronized with the laser pulses and cannot be distinguished from the Stokes photons of interest. 

In the readout step (anti-Stokes scattering), the interaction Hamiltonian coupling the phonon ($c_n$) and anti-Stokes ($b_n$) operators has the form of a beam-splitter interaction, $\lambda b_n^{\dag}c_n+h.c.$. This Hamiltonian implements a state swap between the two modes, with an efficiency dependent on the interaction strength $\lambda$ and duration (here the pulse duration). Therefore, we can model our {readout} scheme by a detector acting directly on the phonon modes $c_n$ but with a reduced detection efficiency $\eta_b=\eta_a\eta_r$ ($\eta_r$ is the {readout} efficiency) and typically larger noise probability $q_b$ (because the {read} pulse is stronger) 

\begin{equation} \label{detb}
D_{\sum_n b_n}(\eta_b,q_b)\doteq D_b^N= \mathbf{1}-(1-q_b)(1-\eta_b)^{\sum_n b_n^{\dag}b_n}
\end{equation}

\vspace{3mm}

\paragraph{Stokes--Stokes auto-correlation ---}

Following the procedure detailed in Ref.~\cite{sekatski2012} we can compute the second-order auto-correlation function of the Stokes modes. To be faithful to the experiment, where a beam-splitter is used to send half of the signal to each detector in this case, we need to divide the detection efficiency by 2. We obtain

\begin{align} \label{gaa}
g^{(a,a)} = \frac{C_{a,a}}{(\tilde S_a)^2} 
\end{align}
with
\begin{align*} 
\tilde S_{a} &=  1-(1-q_a)\left(\frac{(1-\bar p)}{1-\bar p(1-\eta_a/2)}\right)^N 
\end{align*}
\begin{align*} 
C_{a,a}= 2 S_a -1 +(1-q_a)^2\left(\frac{(1-\bar p)}{1-\bar p(1-\eta_a)}\right)^N
\end{align*}

\vspace{3mm}

\paragraph{Witnessing monomode photon pair emission ---}

\begin{figure}[h!]
\centering
\includegraphics[scale=0.9]{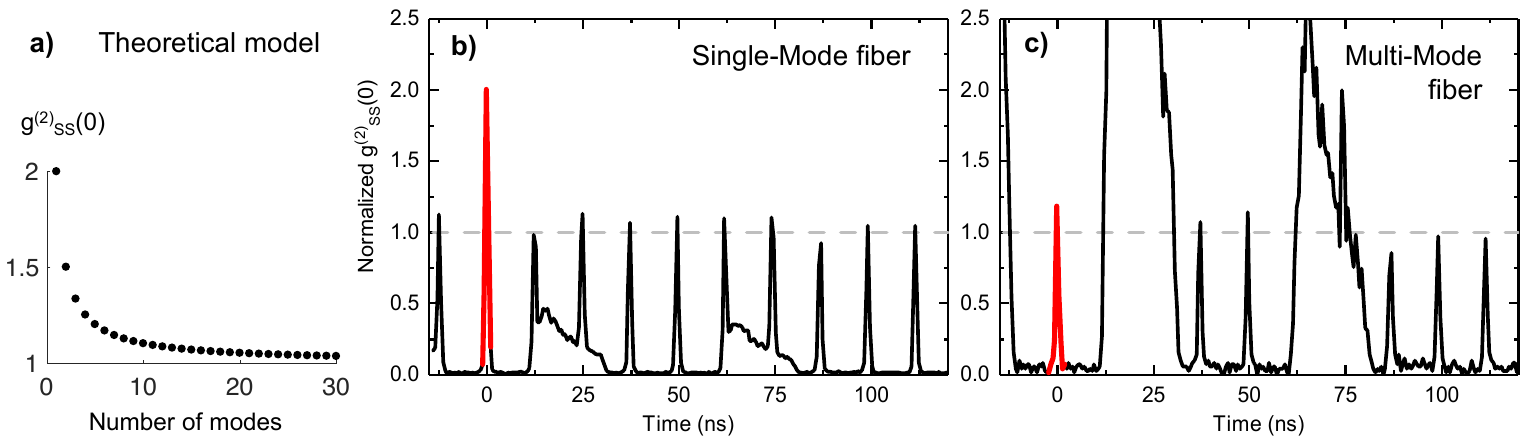}
\caption{\label{fig:Nmode} (a) Predicted Stokes-Stokes auto-correlation as a function of the number of modes $N$ (full expression \eqref{gaa}). The parameters are: $\bar{p}=10^{-3}$, $q_a=q_b=0$ and $\eta_a = \eta_b = 1$, but the results are insensitive on the detection efficiency and noise for a broad range of parameters (typically as long as $q_a,q_b\ll \bar{p}$). (b-c)
Measured Stokes-Stokes auto-correlation function at zero time delay, $g^{(2)}_{S,S}(0)$, using a fiber beam-splitter instead of a dichroic mirror to split the Stokes signal between the two detectors. In (b) the signal is first coupled into a single mode fiber (as in the main text), but not in (c), where it directly couples into the multimode fibers of the detectors. The two broad features on each side of the central peak are due to cross-talk between the detectors, caused by broadband photon emission during an avalanche. When one detector is activated, some of these photons are reflected on the optics and fiber ends into the other detector. (This cannot occur in the $g^{(2)}_{S,aS}$ measurement thanks to the non-overlapping band-pass filters placed before each detector.) Thanks to the sufficiently long optical path, the cross-talk peaks are well separated and do not prevent from measuring $g^{(2)}_{S,S}(0)$ accurately.
}\end{figure}

Because we use a bulk diamond crystal, there exists a continuum of possible optical phonon wavevectors $k$. The flat phonon dispersion around the $\Gamma$ point (the region accessible in Raman scattering lies within the light cone, $k\simeq0$) means that all these phonon modes are degenerate. During inelastic Raman scattering, wavevector and energy are conserved. There is therefore a one-to-one correspondence between the spatial mode of a detected Stokes photon and the spatial mode of the created phonon. By using a single mode fiber to collect the Raman signal in the experiment, our aim is to probe a single phonon mode. We can now use the model and compare its predictions to the Stokes auto-correlation measurements in order to verify the number of phonon modes involved. 

For a fair comparison between the monomode and multimode cases, we first set $\bar p = \frac{p}{N-p(N-1)}$ in Eq.~\eqref{gaa},  that is, the average photon number, given by $\frac{p}{1-p},$ is made independent of the number of modes. We then take the Taylor series in $p$ and $q_a.$ At first order, this gives
\begin{equation}
g^{(a,a)}=1+\frac{1}{N}.
\end{equation}
In the single mode case, it is well known that after tracing out the mode $c$, mode $a$ is in a thermal state, featuring $g^{(a,a)}=2$ at zero delay. Yet, the bunching behavior is lost as $1/N$ when $N$ modes participate the the collected signal, see Fig.~\ref{fig:Nmode}a. Comparison with our experimental results, taken with either a single or multi-mode fiber in collection (Fig.~\ref{fig:Nmode}b,c), confirms that a single phonon mode is probed in the conditions corresponding to all data shown in the main text.

\vspace{3mm}

\paragraph{Modeling of the cross-correlation as a function of the write power ---}

We now compute the Stokes and anti-Stokes single detection probabilities, $S_a$ resp. $S_b$, as well as the coincidence probability $C_{a,b}$, for a configuration where each mode is directed to one detector, as in the main text:

\begin{align*} 
S_{a} &= \Tr D_a^N \rho_{ab}^N = 1-(1-q_a)\left(\frac{(1-\bar p)}{1-\bar p(1-\eta_a)}\right)^N 
\end{align*}

\begin{align*} 
S_{b} &= \Tr D_b^N \rho_{ab}^N = 1-(1-q_b)\left(\frac{(1-\bar p)}{1-\bar p(1-\eta_b)}\right)^N
\end{align*}

\begin{align*} 
C_{a,b}=  \Tr D_a^N D_b^N \rho_{ab}^N = S_a+S_b-1+(1-q_a)(1-q_b)\left(\frac{(1-\bar p)}{1-\bar p(1-\eta_a)(1-\eta_b)}\right)^N
\end{align*}

From this we obtain the normalized second-order cross-correlation between Stokes and anti-Stokes (at a time delay much shorter than the phonon lifetime):

\begin{align} \label{gab}
g^{(a,b)} = \frac{C_{a,b}}{S_a S_b} 
\end{align}

To produce the lines in Fig.~3(c) of the main text, we used formula (\ref{gab}) for a single-mode ($N=1$). The Stokes emission probability is computed by dividing the measured count rate on the Stokes channel by the detection efficiency $\eta$ and the repetition rate. The noise on the Stokes detector is given by the electronic dark counts, and on the anti-Stokes detector it is found by switching off the {write} pulse and measuring the count rate. In this way, all noise sources (in particular thermal phonons and SFWM) are accounted for. Finally, the {readout} efficiency $\eta_r$ is estimated relative to the Stokes emission probability in the \textit{read} pulse $p_{S,r}$ by writing $\eta_r=\alpha_r p_{S,r}$. Indeed, assuming that one phonon has been created, then in the ideal case ($\alpha = 1$) the probability of anti-Stokes scattering (i.e. {readout} of this phonon) should be equal to the probability of Stokes scattering from the vacuum state.
With a detection efficiency $\eta_a=7 \%$, close to the value of $\sim 10\%$ estimated by measuring the transmission of the laser beams and multiplying by the manufacturer's value of the SPAD efficiency, and a readout efficiency $\alpha_r = 30 \%$, we obtain a good agreement with the experimental data points of Fig.~3 taken at three different {read} powers. 

\vspace{3mm}

\paragraph{Insight into the phononic state ---} 

\begin{figure}[h!]
\centering
\includegraphics[scale=0.9]{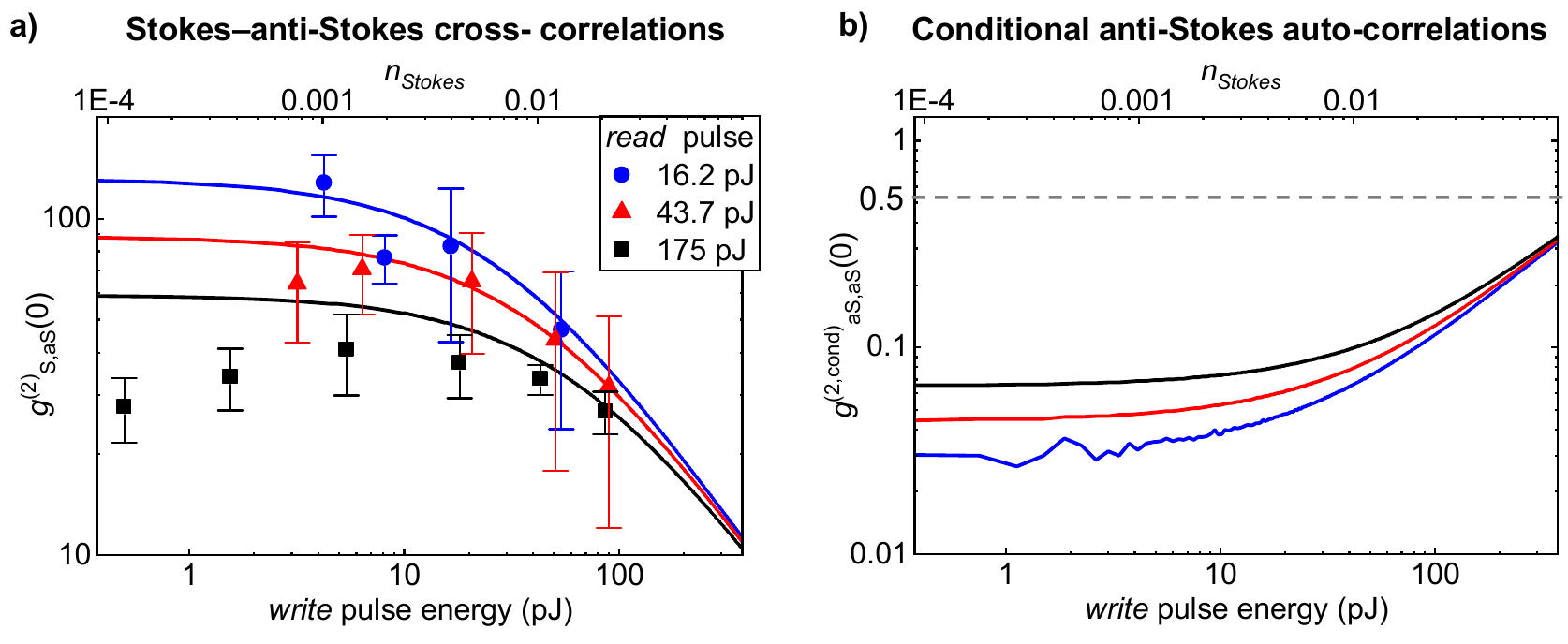}
\caption{\label{fig:g2cond} (a) Reproduction of Fig.~3c of the main text. (b) Inferred value of the anti-Stokes auto-correlation function  conditional on the detection of a Stokes photon in the {write} pulse, determined using eq.~\eqref{cond_g2}, for the same parameters (noise and detector efficiencies) used to reproduce the data in panel (a).
Although the three-fold coincidence measurement is not feasible due to the low experimental aS count rate, our analysis shows that heralded single photons are in principle producible by this approach. 
}\end{figure}

In this paragraph, we show that our experimental results are compatible with a description of the phononic state as a single phonon Fock state. Focusing on a single mode description of the joint phonon--Stokes-photon state, the phononic state conditioned on a Stokes detection is given by
\begin{equation}
\rho_{c|a}=\frac{(1-p)}{\rm{tr}_c  \rho_{c|a}} \sum_{n \geq 0} p^n (1-(1-q_a)(1-\eta_a)^n) |n\rangle\langle n|
\end{equation}
with 
\begin{equation}
{\rm tr}_c  \rho_{c|a} = 1-\frac{(1-q_a)(1-p)}{1-p(1-\eta_a)}
\end{equation}
Swapping this state to the anti-Stokes mode by means of the read pulse and performing subsequently an auto-correlation measurement would lead to
\begin{equation}
\label{cond_g2}
g^{(b,b)}_{|a} = \frac{1-2(1-q_b) \zeta(1-\eta_b/2)+(1-q_b)^2 \zeta(1-\eta_b)}{(1-(1-q_b) \zeta(1-\eta_b/2))^2} 
\end{equation}
where 
\begin{equation}
\zeta(x)=\frac{1}{\rm{tr}_c  \rho_{c|a}} \left(\frac{1-p}{1-p x}-\frac{(1-q_a)(1-p)}{1-p(1-\eta_a)x}\right).
\end{equation}

Using the set of parameters that we found to fit the result of the cross-correlation measurement as a function of the write power (Fig. 3c of the main text), we plot the result that would be obtained by performing an auto-correlation measurement of the conditional anti-Stokes field as a function of the write power using Eq. \eqref{cond_g2}, see Fig.~\ref{fig:g2cond}. Most of our data are in the parameter regime where $g^{(b,b)}_{|a}\ll 1/2,$, which well justifies a description of the conditional phonon state by a single phonon Fock state.


\section{Full numerical quantum model: Master equation approach}\label{sec:numeric}

In this section we present an effective Hamiltonian faithfully modeling the dynamics of the experiment and the non-zero temperature of the phonon bath. We then compute the second-order S--aS cross-correlation function assuming ideal single-photon detectors.
We build on the work of \cite{parra-murillo2016}, in which a single laser pulse is considered, and add one mode for the Stokes photon generated by the first laser pulse. Our model thus takes into account one phononic mode, the Stokes photons produced by the first laser pulse $S_1$ and both Stokes and anti-Stokes photons produced by the second laser pulse ($S_2$ respectively $aS_2$). 
The time dependent excitation sources (laser pulses) are treated as classical coherent fields. After transforming the photonic operators to the frame rotating at $\omega_{0}=(\omega_{L_1}+\omega_{L_2})/2$, the Hamiltonian is given by (in units where $\hbar=1$)
\begin{align}
	\begin{split}
		\mathcal{H} =& \Delta_{S_1}\co[a]_{S_1}\ao[a]_{S_1}+\Delta_{S_2}\co[a]_{S_2}\ao[a]_{S_2}+\Delta_{aS_2}\co[a]_{aS_2}\ao[a]_{aS_2}\\
		&+ \omega_m\co[b]\ao[b]+\mathcal{H}_{S_1}+\mathcal{H}_{S_2}+\mathcal{H}_{aS_2}
	\end{split}
\label{Hamiltonian S-aS}
\end{align}
where
$\Delta_{x}=\omega_{x}-\omega_0$, $x=S_1,S_2,aS_2$. The different contributions to the Hamiltonian read
\begin{align*}
	\begin{split}
		\mathcal{H}_{S_1}&=
			\lambda_{S_1}\left(\alpha_{L_1}\co[b]\co[a]_S+\alpha_{L_1}^*\ao[b]\ao[a]_S\right)\\
		\mathcal{H}_{aS_2}&=\lambda_{aS_2}\left(\alpha_{L_2}\ao[b]\co[a]_{aS_2}+\alpha_{L_2}^*\co[b]\ao[a]_{aS_2}\right)\\
		\mathcal{H}_{S_2}&=
		\lambda_{S_2}\left(\alpha_{L_2}\co[b]\co[a]_{S_2}+\alpha_{L_2}^*\ao[b]\ao[a]_{S_2}\right)\\
	\end{split}
\end{align*}
$\ao[b](\co[b])$, $\ao[a]_{S_j}(\co[a]_{S_j})$ and $\ao[a]_{aS_j}(\co[a]_{aS_j})$ are respectively the annihilation (creation) operator of the phonon, Stokes and anti-Stokes photons produced by laser $j$. The $\lambda_x$ are the coupling constants.
The laser fields are represented by the complex numbers $\alpha_{L_k}(t)$ with a Gaussian envelope
\begin{equation*}
\alpha_{L_k}(t)=
A_{L_k} \exp\left(-(t-t_{0_{L_k}})^2/(2\sigma_{t,L_k}^2)\right) \exp(-i\Delta_{L_k} t)
\end{equation*}
 where $A_{L_k}$, $t_{0,L_k}$, $\sigma_{t,L_k}$ and $\Delta_{L_k}=\omega_{L_k}-\omega_0$ are the excitation amplitude, time, bandwidth and relative frequency of the pump $L_k$. We choose real values for $A_{L_k}$ since our scheme is not phase sensitive. 
A Markovian master equation approach is used to compute the dynamics of the open quantum system. It governs the time evolution of the reduced density matrix
$\op{\rho}=Tr_{R} (\op{\rho}_{tot})$
when tracing out the degrees of freedom of the reservoir. 
In the Lindbladian form, it reads
\begin{align}
\dfrac{d\op{\rho}(t)}{dt}=&\dfrac{1}{i\hbar}\commut{\mathcal{H}}{\op{\rho}} +\Lindbladian{\rho}\\
\Lindbladian{\op{\rho}}=&\Lindbladian{\op{\rho}}{mech}+\Lindbladian{\op{\rho}}{S_1}+\Lindbladian{\op{\rho}}{S_2}+\Lindbladian{\op{\rho}}{aS_2}
\end{align}
where
$\Lindbladian{\op{\rho}}{x}=\gamma_x\left(\bar{n}_{th,x}+1\right)\dissip{\ao}\op{\rho}+\gamma_x\bar{n}_{th,x}\dissip{\co}\op{\rho}$. 
The phonon bath thermal occupancy $\bar{n}_{th,mech} \simeq 1.7\pw{-3}$ at 300~K. The occupancy of the photonic baths are chosen to reproduce the noise in our experiment (see below).
The decay rates $\gamma_x=2\pi/\tau_x$  are computed using the parameters $\tau_x$, given by the pulse durations for the photons and by the phonon lifetime. The standard dissipator is 
$$\Dissip{\op{A}}{\op{B}}.$$

To analyze the S--aS correlation, we compute the two times second order correlation function


\begin{equation*}
g_{S,aS}^{(2)}(t_1,t_2)=
\dfrac{\mean{\co[a]_S(t_1)\co[a]_{aS}(t_2)\ao[a]_{aS}(t_2)\ao[a]_S(t_1)}}
{\op{n}_{S}(t_1)\op{n}_{aS}(t_2)}
\end{equation*}
and use the formula \cite{flayac2015} 
\[
\mean{\op{A}(t_1)\op{B}(t_2)\op{C}(t_1)}=Tr\left(\op{B} U(t_1,t_2)\op{C}\op{\rho}(t_1)\op{A}\right)
\]
where $\op{A}(t_1)=\co_{S}(t_1)$, $\op{B}(t_2)=\co_{aS}(t_2)\ao_{aS}(t_2)$, $\op{C}(t_1)=\co_S(t_1)$, and $U(t_1,t_2)$ is the evolution operator from $t_1$ to $t_2$.

This assumes that the single photon detectors are ideal (they have no noise and $100\%$ efficiency) and spectral filtering is ideal too, meaning that each detector is only sensitive to the $S_1$, resp. $aS_2$ mode. In the next section, we explain how we include noise in the model, but it remains idealized with respect to detection efficiency and filtering, which is why we don't expect quantitative agreement with the experiment. 
 
There is no fitting parameter in this model. All frequencies are fixed by the experimental settings. Furthermore, we choose $\lambda_{S_1}=\lambda_{S_2}=\lambda_{aS_2}$, since only the products $\lambda A_{L_k}$ can be compared to the experiment through the photon populations that are generated in the two pulses, and which are directly measured, \textit{modulo} our detection efficiency. The last free parameter is the phonon lifetime, which is set to 4~ps to reproduce the observed S--aS correlation decay time. Finally, as detailed in Section~\ref{sec:back}, we take several noise sources into account: detector dark counts, spontaneous four wave mixing, and SaS pair emission from a single pulse. 

\begin{figure}[h!]
\centering
\includegraphics[width=14cm]{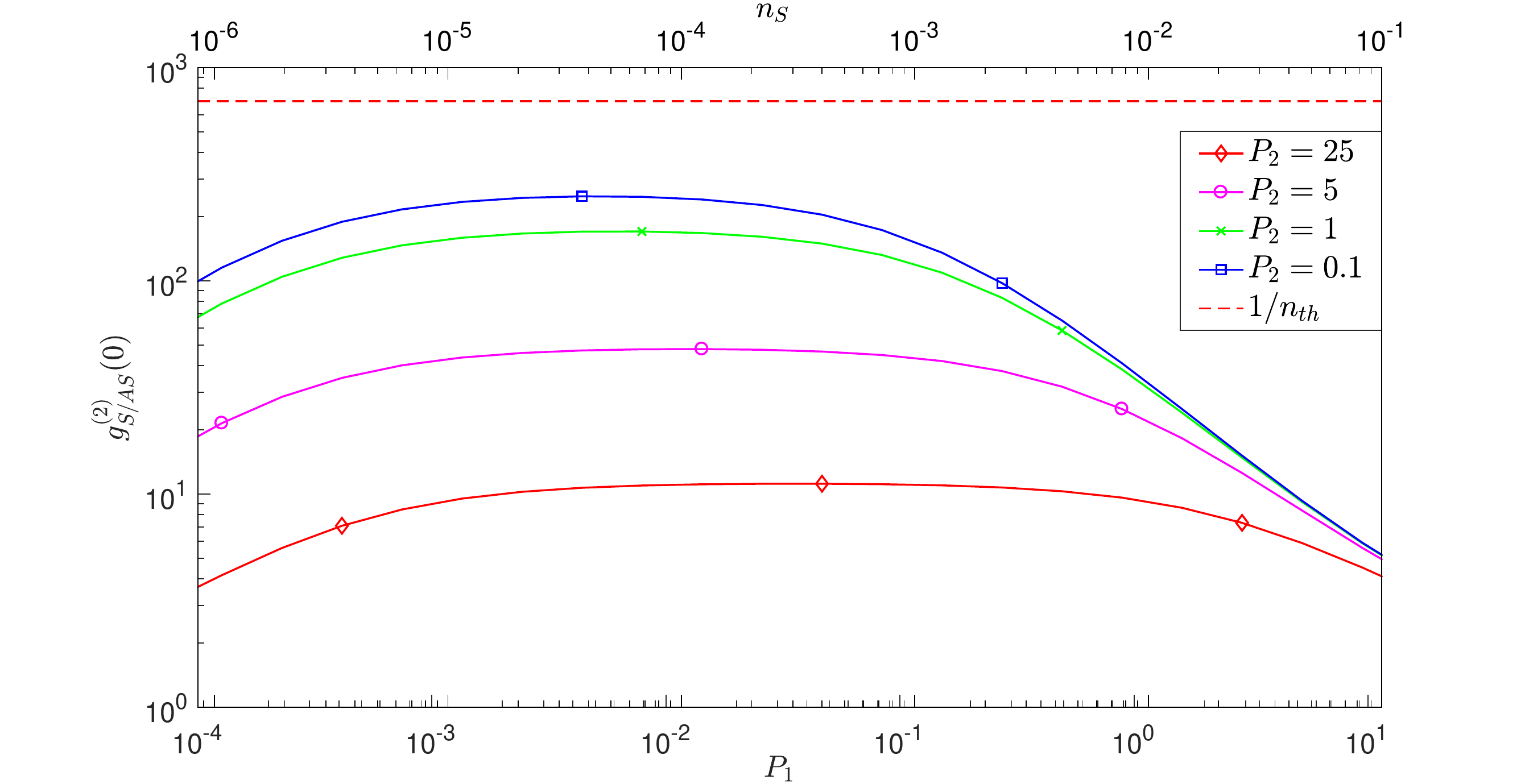}
\caption{Numerical results of the full, non-zero temperature quantum model. Values of $g_{S,aS}^{(2)}(0)$ as a function of \modif{the write pulse energy $P_1$ (arbitrary units, lower axis) and the computed resulting Stokes mode occupancy generated by Raman scattering in the {write} pulse (upper axis), plotted for different values of the {read} pulse energy $P_2$ (same units as write pulse). The red dashed curve is the constant $1/n_{th}$.}
The parameters used in the calculation are given in the SM.}
\label{fig:numeric}
\end{figure}

The results of the numerical resolution of the model are shown in Fig.~\ref{fig:numeric} and are in good qualitative agreement with our measurements (Fig.~3c in the main text). They show that the magnitude of the correlation drops as $1/n_S$ at high {write} power, so that higher correlations are achieved by reducing the {write} pump power, thereby reducing the probability of two-phonon emission. At the lowest {write} power, the influence of dark counts becomes non-negligible and the correlation drops. The numerical results also confirm that our understanding of the noise sources (see below) explains qualitatively well the behavior of the maximum achievable  $g_{S,aS}^{(2)}(0)$ as a function of the {read} power. 

Importantly, the numerical model shows that even without any other noise source than the Raman process, SaS pair emission in the {read} pulse does limit the achievable $g_{S,aS}^{(2)}$ below the value of $1/n_{th}$ expected from the thermal phonons background. 

\modif{Finally, we note that in Fig.~3c of the main text, for the data points at the highest read power (black squares), a decrease of $g_{S,aS}^{(2)}(0)$ for decreasing write power may be observed (although the changes are within the error bars). Intriguingly, this occurs for mean Stokes photon numbers well above the dark counts level, where Fig.~\ref{fig:numeric} predicts a plateau. Possible explanation for this discrepancy, apart from the idealized 100\% efficiency of the detectors, is the existence of other sources of noise not taken into account in the model; they include fluorescence from the sample and the optical elements of the setup, and residual leakage of laser light onto the detectors.} 

The parameters used in the calculation of Fig.~\ref{fig:numeric} are summarized below:

\begin{table}[h!]
\label{table:table_para_sim}
\begin{tabular}{|c|c|}
             \hline
             $\sigma_{t,L_k}$&$0.2 \text{ ps}$\\
             \hline
             $\omega_{L_1}$&$2.90\cdot 10^{3} \text{ THz}$\\
             \hline
             $\omega_{L_2}$&$2.35\cdot 10^{3} \text{ THz}$\\
             \hline
             $\omega_{m}$&$2\pi\cdot 40 \text{ THz}$\\
             \hline
             $\tau_x$&$\sigma_{t,L_k}$\\
             \hline
             $\tau_m$&$4 \text{ ps}$\\
             \hline
    $T$&$300 \text{ K}$\\
    \hline
\end{tabular}~~\\
\caption{Parameters used in the numerical calculation of Fig.~5.}
\end{table}
where $\tau_x=2\pi/\gamma_x$ and $x=S_1,S_2,aS_2$.


\section{Investigation of the sources of noise and their modeling}\label{sec:back}

\begin{figure}[h!]
\centering
\includegraphics[width=16cm]{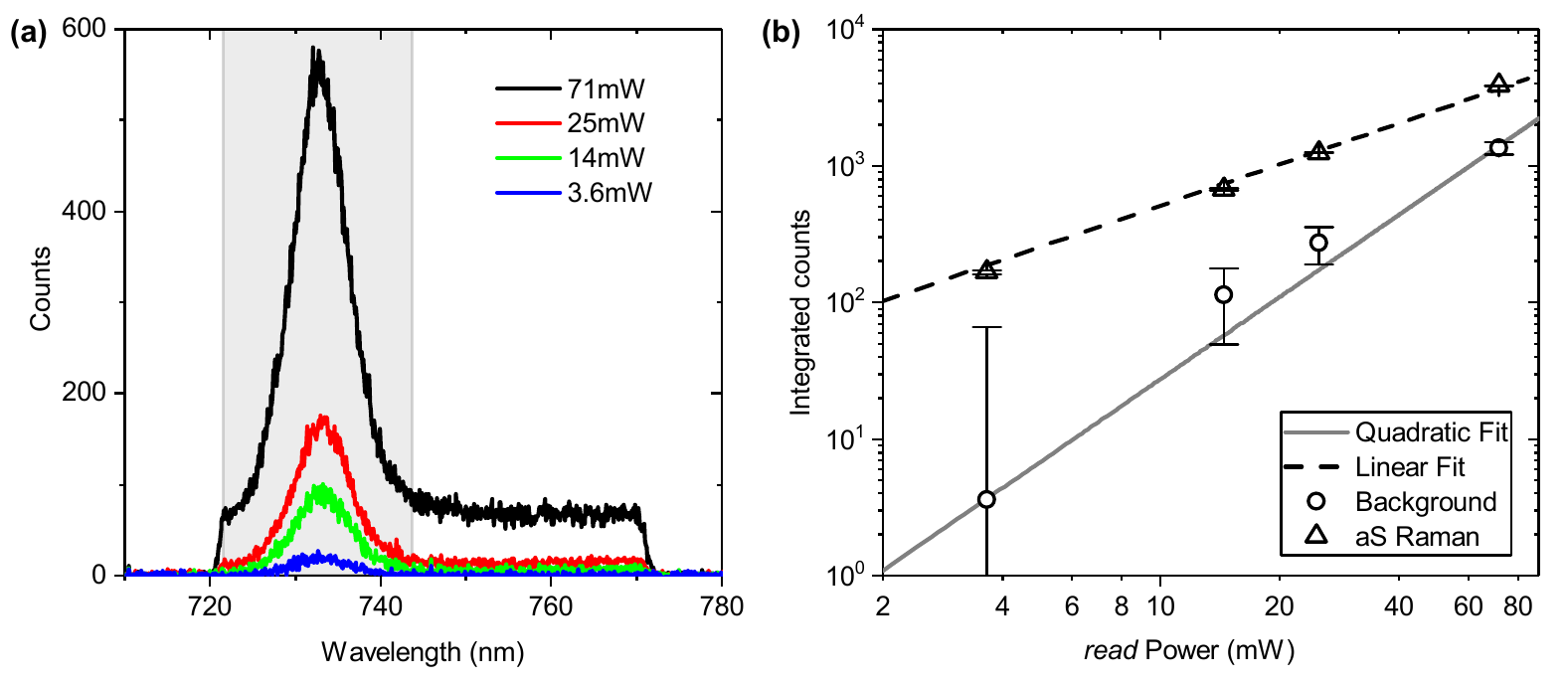}
\caption{\label{fig:aSBackground} (a) Anti-Stokes spectra generated by the {read} pulse at different powers ({write} pulse is off). We can see a spectrally flat background which overlaps with the aS detection window marked in light gray (b) Anti-Stokes and background areas (within the aS detection window) as a function of laser power, together with a linear and quadratic fit, respectively.
}\end{figure}

An intrinsic mechanism in diamond for generating noise photons from a single pulse is spontaneous degenerate four-wave mixing (SFWM). In this nonlinear $\chi_3$ process two photons from the laser pulse are annihilated to generate two photons at higher and lower energies. Because of the small refractive index dispersion in diamond around our wavelengths, energy and momentum conservations are satisfied for any pair of photons symmetrically spaced in energy around the pump. This leads to a spectrally flat background. 
While this process is much weaker than spontaneous Stokes scattering, we found that it was non-negligible compared to anti-Stokes scattering in our experimental conditions. 

We verified this hypothesis by measuring the dependence of the anti-Stokes peak and background noise on the power of the Ti:Sapph beam, Fig.~\ref{fig:aSBackground}. As expected for a two-photon process, the background scales quadratically with power. The anti-Stokes signal scales linearly, which means that we are in a regime were thermally excited phonons are the dominating source of anti-Stokes scattering. 
Since the SFWM photons generated by the {read} pulse are not correlated with the Stokes photons generated by the {write} pulse, they contribute to accidental coincidences (side peaks in the coincidence histograms) and to a decrease of the measured S--aS correlation.

\vspace{3mm}
\paragraph{Theoretical modeling of FWM noise ---}

We model the SFWM noise generated in each pulse by setting an effective temperature for \modif{both} the Stokes and anti-Stokes modes that instantaneously follows the temporal profile of the laser pulses. \modif{The noise generated from each pulse is felt by both detectors because it is spectrally broad. However, a photon pair generated in SFWM cannot trigger coincidences on the two detectors due to the mismatched configuration of the spectral detection windows for this process.}

Since the process involves 2 pump photons, the noise temperature depends quadratically on the laser power:
\begin{equation}
n_{th,noise,k} = n_{th,0}\left(c_2 A_{L_k}(t)^4+c_1\right),
\end{equation}
where
\begin{equation}
A_{L_k}(t)=A_{L_k} \exp\left(-(t-t_{0_{L_k}})^2/(2\sigma_{t,L_k}^2)\right)
\end{equation}
is the temporal shape of the pulse amplitude and
 $k=1(2)$ for the {write (read)} pulse. The constant $c_1$ accounts for the electronic dark count probability of the detectors $c_1=\kappa_d \tau_{bin} = 10^{-6}$ with $\kappa_d \simeq 500$~Hz the dark count rate and $\tau_{bin} \simeq 2$~ns the histogram binning time. This contribution to the noise is negligible in most experimental conditions. 
Based on the preceding analysis, we could deduce that for a total aS emission probability of $n_{aS,tot}=2.4\cdot 10^{-4}$ (under excitation with only the {read} pulse), the contribution of SFWM noise was $n_{th,aS} = 0.6\cdot 10^{-4}$. This measurement was used to set the value of $c_2=4.5\cdot 10^{-6}$ in order to account for SFWM noise in the model of Sec.~\ref{sec:numeric}. 

\vspace{3mm}

\paragraph{Noise from simultaneous SaS pairs ---}

Another source of uncorrelated aS photons in the second pulse is the process in which two photons of the {read} pulse interact via the exchange of the same phonon to produce a SaS pair. To account for this effect in the model, we had to include the Stokes mode in the {read} pulse, although it is not detected in the experiment. Including the anti-Stokes mode in the {write} pulse did not measurably modify the results, so it was not included in the final model.
The main consequence of this process is that the correlations drop at high {read} power, even in the absence of SFWM noise. One way to mitigate this effect in future work would be to create an asymmetry between the S and aS emission rate in the second pulse, for example using a cavity.


\section{$g_{S,aS}^{(2)}(\tau)$ calculation from correlation measurements}

In order to estimate the normalized correlation function $g_{S,aS}^{(2)}(\tau)$, we used a binning time of 1.536 ns around each coincidence peak. After subtracting the flat, totally uncorrelated background due to the dark counts (generally very small), the average counts in the side peaks is computed, and $g_{S,aS}^{(2)}(\tau)$ is the ratio of the central peak (0~ns) to the side peaks. The histograms shown in the main text are raw data, before background subtraction and binning (acquisition bin size = 512~ps).

The uncertainty in the area of the side peaks is taken as the standard deviation of the first 25 peaks after the zero delay peak. This is then propagated as  
\begin{equation}
\left(\delta g_{S,aS}^{(2)}(\tau)\right)^{2} = \left(g_{S,aS}^{(2)}(\tau)\right)^{2} \left[ \left(\frac{\delta A_{SP}}{A_{SP}}\right)^{2} + \left(\frac{\delta A_{CP}}{A_{CP}}\right)^{2}\right]
\end{equation}
where $\delta g_{S,aS}^{(2)}(\tau)$ is the uncertainty in $g_{S,aS}^{(2)}(\tau)$, $A_{SP}$ is the average area of the side peaks, $\delta A_{SP}$ is the standard deviation in the side peak area, $A_{CP}$ is the average area of the correlated peak, and $\delta A_{CP}$ is the standard deviation in the correlated peak area. The random fluctuations follow Poissonian statistics, so we take $\delta A_{CP}$ to be $\sqrt{A_{CP}}$, leading to
\begin{equation}
\delta g_{S,aS}^{(2)}(\tau) = g_{S,aS}^{(2)}(\tau) \sqrt{ \left(\frac{\delta A_{SP}}{A_{SP}}\right)^{2} + \frac{1}{A_{CP}} }
\end{equation}

\begin{figure}[h!]
\centering
\includegraphics[width=8cm]{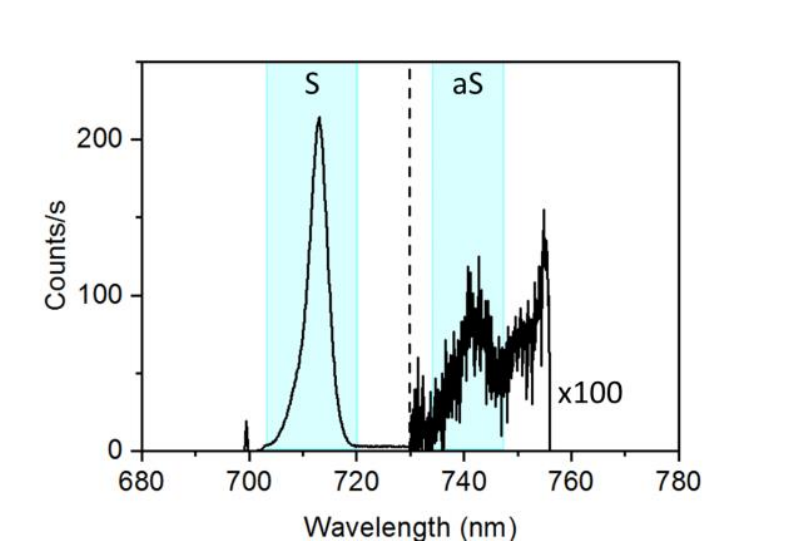}
\caption{\label{fig:ExpSpectra} Spectra showing the Stokes and anti-Stokes peaks under the the experimental conditions for the correlation measurements of Fig.~4 (open circles). In the aS detection window, a non-negligible fraction of the signal is generated by the {write} pulse and leads to accidental coincidences (cross-talk between channels)
}\end{figure}

At certain powers and wavelength configurations, a non-negligible portion of the side peaks' counts comes only from the {write} pulse (see e.g. Fig.~\ref{fig:ExpSpectra}). This happens when the emission spectrum from the {write} pulse has a non-negligible overlap with the aS detection window, most likely due to residual fluorescence from the diamond or the optical parts. When this is the case, a S--aS coincidence histogram is measured with only the {write} pulse, the {read} pulse being blocked, and is subtracted from the histogram obtained with both pulses on. This was done only for the data corresponding to the black curve (open circles) in Fig.~4, and for the points at {write} powers $P_w$ over 1~mW in Fig.~3  of the main text. For all other data points this was not necessary. 


\section{Fitting of delay dependence curves}

The curves for the dependence of $g_{S,aS}^{(2)}(\tau)$ are fit with the following function, which represents the convolution of an exponential decay function with a Gaussian function
\begin{equation}
C+\frac{1}{2}Ae^{\frac{\sigma^{2}}{2\tau^{2}}}e^{-\frac{t-t_{0}}{\tau}}\left(1+\text{erf}\left(\frac{t-t_{0}-\frac{\sigma^{2}}{\tau}}{\sqrt{2}\sigma}\right)\right)
\end{equation}

Where $C$ is a constant representing the value for $g_{S,aS}^{(2)}(\tau \rightarrow \infty)$, $A$ is a constant that depends on the maximum value of the correlation at $\tau=0$, $\sigma$ is the standard deviation of the instrument response function, $t_0$ is the the time at which the pulses overlap, and $\tau$ is the time constant of the exponential decay.

In order to fit the data we set $\sigma=0.22$ ps, as explained in Sec.~\ref{sec:CARS}, and $C=1$, as $g_{S,aS}^{(2)}(\tau)$ will drops down to 1 when the {read} pulse arrives before the {write} pulse, or for sufficiently long delays. The remaining parameters are fitted to the data using the least squares algorithm built in Matlab. Since $t_0$ is only a shift in the time axis, we can consider that our fit has only two free parameters, given by the maximum of the correlation $g_{S,aS}^{(2)}(0)$ and the time constant of its decay. 


\section{CARS signal}\label{sec:CARS} 
\begin{figure}[h!]
\centering
\includegraphics[width=16cm]{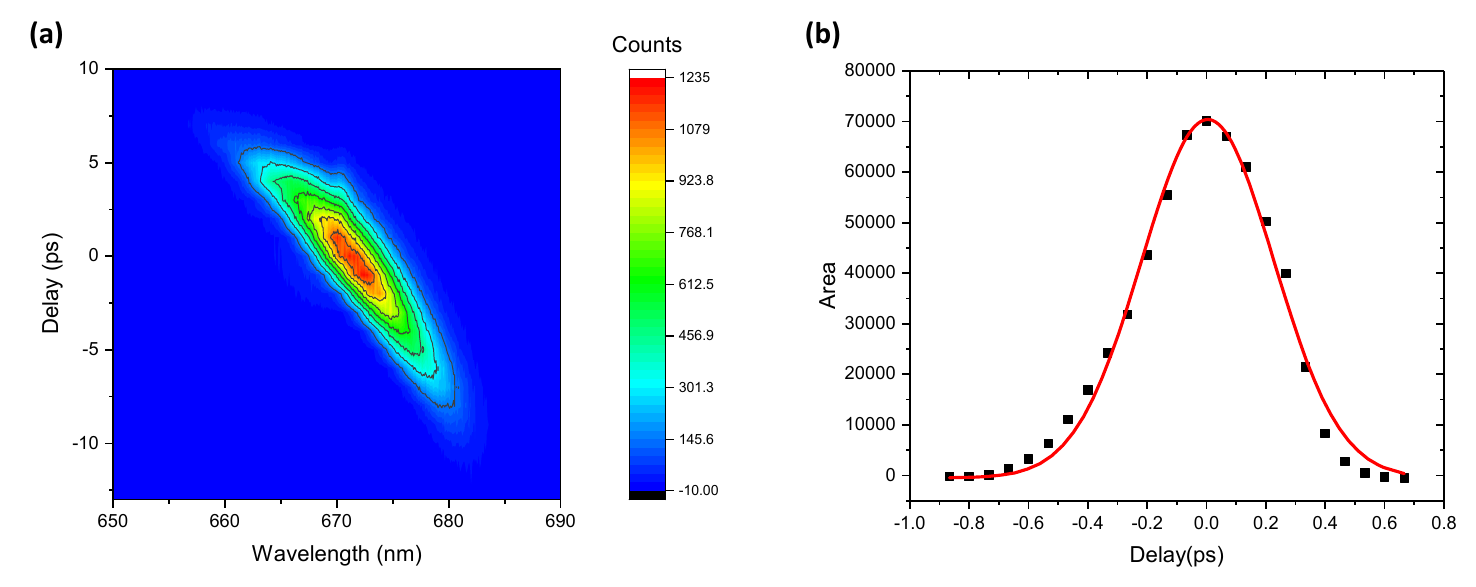}
\caption{\label{fig:CARS} (a) Spectrally resolved Coherent Anti-Stokes Raman Scattering (CARS) as a function of time delay between the {write} and {read} pulses. We interpret the diagonal shape of the signal as result of chirp in one or both pulses.  (b) Wavelength-integrated CARS signal as a function of time delay between the {write} and {read} pulses. We use a Gaussian fit to this data to model the instrument response function, obtaining a standard deviation $\sigma = 220$ fs.
}\end{figure}

Coherent Anti-Stokes Raman Scattering (CARS) was used to overlap the two pulses temporally and characterize our temporal resolution. CARS occurs whenever the frequency difference between the {write} and {read} pulses matches the phonon frequency, $\omega_1-\omega_2=\Omega_m$. Under this condition, one photon from the {write} and one photon from the {read} pulse stimulate the emission of one phonon, which interacts with a second photon from the {write} pulse to generate a photon at the anti-Stokes frequency $\omega_1+\Omega_m$. The results of this characterization are shown in Fig.~\ref{fig:CARS}.

\end{document}